\documentclass[twocolumn,showpacs,preprintnumbers]{revtex4}%
\usepackage{amsfonts}
\usepackage{amssymb}
\usepackage{amsmath}
\usepackage{graphicx}
\usepackage{dcolumn}
\usepackage{bm}
\usepackage[toc]{appendix}
\usepackage{array}
\usepackage{hyperref}
\usepackage{lipsum}%
\providecommand{\U}[1]{\protect\rule{.1in}{.1in}}
\graphicspath{{figures/}}
\begin{document}
\title{Phase sensitivity at the Heisenberg limit in an SU(1,1) interferometer via parity detection}
\author{Dong Li$^{1,2}$, Bryan T. Gard$^{2}$, Yang Gao$^{3}$, Chun-Hua Yuan$^{1}$}
\email{chyuan@phy.ecnu.edu.cn}
\author{Weiping Zhang$^{1}$, Hwang Lee$^{2}$}
\email{hwlee@phys.lsu.edu}
\author{Jonathan P. Dowling$^{2}$}

\affiliation{$^{1}$Quantum Institute for Light and Atoms, Department of Physics, East China Normal University, Shanghai 200062, P. R. China\\
$^{2}$Hearne Institute for Theoretical Physics and Department of Physics and
Astronomy, Louisiana State University, Baton Rouge, LA 70803, USA\\
$^{3}$Department of Physics, Xinyang Normal University, Xinyang, Henan 464000,
P. R. China}

\date{\today }

\begin{abstract}
We theoretically investigate the phase sensitivity with parity detection on an
SU(1,1) interferometer with a coherent state combined with a squeezed vacuum state.
This interferometer is formed with two parametric amplifiers for beam
splitting and recombination instead of beam splitters. We show that the
sensitivity of estimation phase approaches Heisenberg limit and give the
corresponding optimal condition. Moreover, we derive the
quantum Cram\'er-Rao bound of the SU(1,1) interferometer.

\end{abstract}

\maketitle

\section{Introduction}

High precision metrology has recently been receiving a lot of attention
\cite{Helstrom,Braunstein,Lee2002,Giovannetti} due to the benefits to
advanced science and technology. One common tool for high precision
measurement is optical Mach-Zehnder interferometry, which typically contains two beam splitters (BS).
Usually coherent light is split by the first BS, then one beam
experiences a phase shift $\phi$ while the other is retained as a reference, and
the two beams combine by a second BS. One can detect the output light to obtain
the phase shift information. However the phase sensitivity, $\Delta\phi$, is
limited by the Shot Noise limit (SNL), $1/\sqrt{N}$ ($N$ is the mean photon
number). This limit is due to the classical nature of the coherent state and can
be surpassed by using nonclassical states of light, such as squeezed states
\cite{Caves} and N00N states \cite{Dowling}. With the help of the nonclassical
states the phase sensitivity can achieve Heisenberg limit (HL) $1/N$.

\begin{figure}[ptb]
\centerline{\includegraphics[scale=0.65,angle=0]{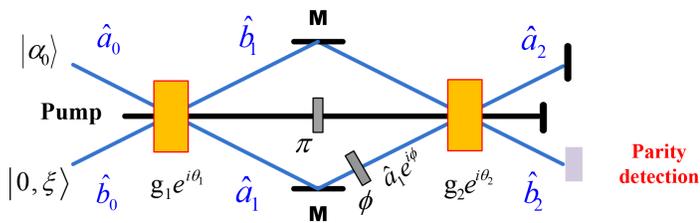}}\caption{(Color
online)\label{fig1} The schematic diagram of SU(1,1) interferometer. Two OPAs take the
place of two beam splitters in the traditional Mach-Zehnder interferometer.
$g_{1}$ ($g_{2}$) and $\theta_{1}$ ($\theta_{2}$) describe the strength and
phase shift in the OPA process $1$ ($2$), respectively. $a_{i}$ and $b_{i}$
($i=0,1,2$) denote two light beams in the different processes. The pump field
between the two OPAs has a $\pi$ phase difference. $\phi$: phase shift; $M$:
mirrors.}%
\end{figure}

Another possibility for beating the SNL is to use an interferometer in which
the mixing of the optical beams is done through a nonlinear transformation,
such as the SU(1,1) interferometer as shown in Fig. \ref{fig1}. This type of
interferometer, first proposed by Yurke \emph{et al}. \cite{Yurke}, is described by
the group SU(1,1), as opposed to SU(2), where nonlinear transformations are
optical parametric amplifiers (OPA) or four-wave mixers. Yurke \emph{et al}.
\cite{Yurke} pointed out that this type of interferometer can reach the HL
but only with small photon numbers.

Recently, a new theoretical scheme was proposed to inject strong coherent
light to ``boost" the photon number in an SU(1,1) interferometer with
intensity detection by Plick \emph{et al}. \cite{Plick}. Their scheme circumvents the
small-photon-number problem. Jing \emph{et al}. \cite{Jing} reported the experimental
realization of such an interferometer. In this nonlinear
interferometer, the maximum output intensity can be much higher than the input
due to the parametric amplification. Marino \emph{et al}. \cite{Marino} investigated
the loss effect on phase sensitivity of the SU(1,1) interferometers with intensity
detection. While they show that propagation losses degrade the phase sensitivity, it is
still possible to beat SNL even with a significant mount of loss. Hudelist \emph{et al}. \cite{Hudelist} observed an improvement of 4.1 dB in signal-to-noise ratio
compared with an SU(2) interferometer under the same operation condition.
More recently, Li \emph{et al}. \cite{Li} showed that an SU(1,1) interferometer with coherent and
squeezed input states via homodyne detection can approach the HL.

Heisenberg-limited sensitivity of phase measurement is one goal of quantum
optical metrology. For this purpose, the search for the optimal detection
scheme still continues. Here, we consider parity measurement as our detection
scheme. Parity detection was first proposed by Bollinger \emph{et al}.
\cite{Bollinger} in 1996 to
study spectroscopy with a maximally entangled state of trapped ions. It was later adopted for an optical interferometer by Gerry
\cite{Gerry2000}. Mathematically, parity detection is described by a simple,
single-mode operator $\hat{\Pi}_{a}=(-1)^{\hat{a}^{\dag}\hat{a}}$, which
acts on the mode $a$. Hence, parity is simply the evenness or oddness of the
photon number in an output mode. In experiments, the parity operator can be
implemented by using homodyne techniques \cite{PlickHomo} for high power,
or observing the photon-number distribution with a photon-number resolving
detector for small photon numbers.

Here, we still consider coherent and squeezed vacuum input light as in
our previous work \cite{Li} which focused on homodyne detection on SU(1,1)
interferometers. With the coherent state and squeezed vacuum state as the input, one
can reduce the required intensity of input coherent states and obtain the same
measurement sensitivity, which can eliminate the disadvantages due
to using stronger coherent states. Furthermore, as shown in Ref. \cite{Li},
these input states in an SU(1,1) interferometer are shown to approach the
Heisenberg limit when the mean photon numbers in coherent state and squeezed
vacuum state are roughly equal, with a stronger OPA process (strength $g>2$).
This optimal condition for SU(1,1) interferometers is similar to the SU(2)
case \cite{Seshadreesan,Pezze}. For a Mach-Zehnder Interferometer (MZI) injected by coherent and squeezed vacuum light, the phase
sensitivity with parity detection is $1/\sqrt{\left\vert \alpha\right\vert
^{2}e^{2r}+\sinh^{2}r}$. When a coherent state and
squeezed vacuum state are in roughly equal intensities, $|\alpha|^{2}%
\simeq\sinh^{2}r$, the phase sensitivity reaches the HL.

In this paper, we study parity detection on an SU(1,1) interferometer with
coherent and squeezed-vacuum input states. We found that parity detection
can be regarded as an optimal measurement scheme since it has a slightly better
phase sensitivity than homodyne detection. Due to achieving the HL, we also compared
the phase sensitivity with another quantum limit, the quantum Cram\'er-Rao bound
(QCRB) which sets the
ultimate limit for a set of probabilities that originated from measurements on
a quantum system. The QCRB is asymptotically achieved by the maximum likelihood estimator
and gives a detection-independent phase sensitivity $\Delta\phi_{\text{QCRB}}.$

This paper is organized as follows: In Section II we first present the
propagation of input fields through the SU(1,1) interferometer. Then we
discuss how the phase sensitivity approaches the Heisenberg limit and compare the phase sensitivity with the QCRB in Section III. Last, we conclude with a summary.

\section{Parity detection on an SU(1,1) interferometer}

\subsection{Model}

Fig. \ref{fig1} presents the model of an SU(1,1) interferometer, in which
the OPAs replace the 50-50 beam splitters in a traditional MZI. Here we
consider a coherent light mixed with a squeezed vacuum light as input and
\ $\hat{a}$ $(\hat{a}^{\dag})$ and $\hat{b}$ $(\hat{b}^{\dag})$ are the
annihilation (creation) operators corresponding to the two modes. After the first
OPA, one output is retained as a reference, while the other one experiences a
phase shift. After the beams recombine in the second OPA with the
reference field, the outputs are dependent on the phase difference $\phi$
between the two modes.

Next, we will focus on the evolution of an SU(1,1) interferometer in phase
space. The initial Wigner function of the input state, a product state
$|\alpha_{0}\rangle\otimes|0,\xi=re^{i\theta_{s}}\rangle$, with coherent light
amplitude $\alpha_{0}=|\alpha_{0}|e^{i\theta_{\alpha}}$, is given by%
\begin{equation}
W_{\text{in}}(\alpha_{i},\alpha_{0};\beta_{i},r)=W_{|\alpha_{0}\rangle}(\alpha
_{i},\alpha_{0})W_{|0,\xi\rangle}(\beta_{i},r),
\end{equation}
with the Wigner function of coherent and squeezed vacuum state being
\cite{Walls}%
\begin{align}
W_{|\alpha_{0}\rangle}(\alpha_{i},\alpha_{0})  &  =\dfrac{2}{\pi}%
e^{-2|\alpha_{i}-\alpha_{0}|^{2}},\\
W_{|0,\xi\rangle}(\beta_{i},r)  &  =\frac{2}{\pi}e^{-2|\beta_{i}|^{2}%
\cosh2r+(\beta_{i}^{2}+\beta_{i}^{\ast2})\sinh2r},
\end{align}
where $\theta_{s}$ has been set $0$ by appropriately fixing the irrelevant
absolute phase $\theta_{\alpha}$.

After propagation through the SU(1,1) interferometer the output Wigner
function is written as%
\begin{equation}
W_{\text{out}}\left(  \alpha_{f},\beta_{f}\right)  =W_{\text{in}}[\alpha_{i}(\alpha
_{f},\beta_{f}),\beta_{i}(\alpha_{f},\beta_{f})],
\end{equation}
where the relation between variables is described by%
\begin{equation}
\left(
\begin{array}
[c]{c}%
\alpha_{i}\\
\beta_{i}^{\ast}%
\end{array}
\right)  =\hat{T}^{-1}\left(
\begin{array}
[c]{c}%
\alpha_{f}\\
\beta_{f}^{\ast}%
\end{array}
\right)  ,
\end{equation}
where $\alpha_{i},\beta_{i},\alpha_{f}$ and $\beta_{f}$ are the complex
amplitudes of the beams in the mode $\hat{a}_{0},\hat{b}_{0},\hat{a}_{2}$ and
$\hat{b}_{2}$, respectively and $\beta_{f}^{\ast}$ is the conjugate of
$\beta_{f}$. Generally, propagations through the first OPA, phase shift and second
OPA are described by%
\begin{align}
\hat{T}_{\text{OPA1}} & =\left(
\begin{array}
[c]{cc}%
u_{1} & v_{1}\\
v_{1}^{\ast} & u_{1}%
\end{array}
\right)  ,\\
\hat{T}_{\phi} & =\left(
\begin{array}
[c]{cc}%
e^{i\phi} & 0\\
0 & 1
\end{array}
\right)  ,\\
\hat{T}_{\text{OPA2}} & =\left(
\begin{array}
[c]{cc}%
u_{2} & v_{2}\\
v_{2}^{\ast} & u_{2}%
\end{array}
\right)  ,
\end{align}
with $u_{1}=\cosh g_{1},v_{1}=e^{i\theta_{1}}\sinh g_{1},u_{2}=\cosh g_{2}$
and $v_{2}=e^{i\theta_{2}}\sinh g_{2}$, where $\theta_{1}$ $(\theta_{2})$ and
$g_{1}$ $(g_{2})$ are the phase shift and parametrical strength in the OPA
process 1(2), respectively, see, for example Ref.~\cite{Xu}. Therefore, the SU(1,1) interferometer is described
by $\hat{T}=\hat{T}_{\text{OPA2}}\hat{T}_{\phi}\hat{T}_{\text{OPA1}}$. More specifically, we
assume that the first OPA and the second one have a $\pi$ phase difference
(particularly $\theta_{1}=0$ and $\theta_{2}=\pi$) and same parametrical
strength ($g_{1}=g_{2}=g$). In this case, the second OPA will undo what the
first one does (namely $\hat{a}_{2}=\hat{a}_{0}$ and $\hat{b}_{2}=\hat{b}_{0}$)
when phase shift $\phi=0$, which we call a balanced situation.

The input-output relation of variables has the following form%
\begin{align}
\alpha_{i} & \rightarrow G\alpha_{f}+R\beta_{f}^{\ast},\\
\beta_{i}^{\ast} & \rightarrow-R\alpha_{f}+H\beta_{f}^{\ast},
\end{align}
where $G=A-iB\cosh(2g),H=A+iB\cosh(2g)$ and $R=-iB\sinh(2g)$ with $A=\cos
(\phi/2)e^{-i\phi/2}$ and $B=\sin(\phi/2)e^{-i\phi/2}.$ Therefore, the output
Wigner function of a nonlinear interferometer is described by%
\begin{align}
W_{\text{out}}(\alpha_{f},\beta_{f})=  &  \frac{4}{\pi^{2}}e^{-2\left\vert
G\alpha_{f}+R\beta_{f}^{\ast}-\alpha_{0}\right\vert ^{2}}\nonumber\\
&  \times e^{-2\left\vert -R\alpha_{f}+H\beta_{f}^{\ast}\right\vert ^{2}%
\cosh(2r)}\nonumber\\
&  \times e^{2\operatorname{Re}[(-R\alpha_{f}+H\beta_{f}^{\ast})^{2}%
]\sinh(2r)}. \label{wigner}%
\end{align}

\subsection{Phase sensitivity}

Parity measurement has been proved to be a efficient method of detection in
interferometer for a wide range of input states
\cite{Campos,Gerry2005,Campos2005,Gerry2007}. For many input states, parity
does as well, or nearly as well, as state-specific detection schemes
\cite{Chiruvelli,Gerry}. Furthermore, as has been reported recently, parity
detection with a two-mode, squeezed-vacuum interferometer actually reaches
below the Heisenberg limit, achieving the quantum Cram\'er-Rao bound on phase
sensitivity \cite{Anisimov}.

In this paper, we consider parity detection as our measuring method. The
parity operator detection on output mode $b$ is $\hat{\Pi}_{b}\equiv
(-1)^{\hat{b}_{2}^{\dag}\hat{b}_{2}}.$ From the Wigner function, the parity
operator of $b$ mode is given by \cite{Royer}%
\begin{equation}
\langle\hat{\Pi}_{b}\rangle=\frac{\pi}{2}\int W_{out}(\alpha_{f},0)d^{2}%
\alpha_{f}, \label{signal}%
\end{equation}
In our case, $\langle\hat{\Pi}_{b}\rangle$ is a series of rather complex and
un-illuminating expressions which are shown in Appendix \ref{sec:signal}.

The sensitivity of phase estimation based on the outcome of the parity
detection is estimated as%
\begin{equation}
\Delta\phi_{\text{p}}=\frac{\langle\Delta\hat{\Pi}_{b}\rangle
}{|\frac{\partial\langle\hat{\Pi}_{b}\rangle}{\partial\phi}|},
\end{equation}
which is a ratio of detection noise to the rate at which signal changes as a
function of phase and $\langle\Delta\hat{\Pi}_{b}\rangle\equiv
(\langle\hat{\Pi}_{b}^{2}\rangle-\langle\hat{\Pi}_{b}\rangle^{2})^{1/2}=(1-\langle
\hat{\Pi}_{b}\rangle^{2})^{1/2}.$

The phase sensitivity with parity detection for an SU(1,1) interferometer with
coherent and squeezed vacuum states is found to be minimal at $\phi=0$ and is
given by%
\begin{align}
\Delta\phi_{\text{p}}=  &  \frac{1}{\{N_{\alpha}[\sinh(2r)\cos(2\theta_{\alpha})
+\cosh(2r)]+N_{s}+1\}^{1/2}}\nonumber\\
&  \times\frac{1}{[N_{\text{OPA}}(N_{\text{OPA}}+2)]^{1/2}},
\end{align}
where $N_{\alpha}=|\alpha_{0}|^{2}$ is the intensity of input coherent light,
$N_{s}=\sinh^{2}r$ the intensity of the squeezed vacuum light, and
$N_{\text{OPA}}=2\sinh^{2}g$ the spontaneous photon number emitted from the first OPA
which is related to parametric strength. When $\theta_{\alpha}=0,$ the optimal
phase sensitivity is found to be%
\begin{equation}
\Delta\phi_{\text{p}}=\frac{1}{[(N_{\alpha}e^{2r}+N_{s}+1)N_{\text{OPA}}(N_{\text{OPA}}+2)]^{1/2}},
\label{P}%
\end{equation}
where the subscript p indicates parity detection and the factor $e^{2r}$ results from the input squeezed vacuum beam. If
vacuum input is injected ($N_s = 0 $ and $N_{\alpha} = 0$), the phase sensitivity with parity detection is
reduced to $(\Delta\phi_{\text{p,V}})^{2}=1/[N_{\text{OPA}}(N_{\text{OPA}}+2)]$, which is the same as
result of Yurke's scheme with intensity detection \cite{Yurke}.

\section{Discussion}
\subsection{Heisenberg Limit}

In this section, we compare the optimal phase sensitivity of
parity detection with the HL which is related to the total number of photon
$N_{\text{Tot}}(\equiv\langle\hat{a}_{1}^{\dag}\hat{a}_{1}+\hat{b}_{1}^{\dag}\hat
{b}_{1}\rangle)$ inside the interferometer, not the input photon number as the
traditional MZI. The HL is given by
\begin{equation}
\Delta\phi_{\text{HL}}=\frac{1}{N_{\text{Tot}}}, \label{HL}%
\end{equation}
where the subscript HL represents Heisenberg limit. According to Ref.
\cite{Li} the total inside photon number is%
\begin{equation}
N_{\text{Tot}}=(N_{\text{OPA}}+1)(N_{\alpha}+N_{s})+N_{\text{OPA}}, \label{HL1}%
\end{equation}
where the first term on the right-hand side, $(N_{\text{OPA}}+1)(N_{\alpha}+N_{s})$, results from the amplification
process of the input photon number and the second one is
related to the spontaneous process. Thus the total inside photon number $N_{\text{Tot}}$ corresponds to not only the OPA strength
but also input photon number. 

\begin{figure}[t]
\centerline{\includegraphics[scale=0.9,angle=0]{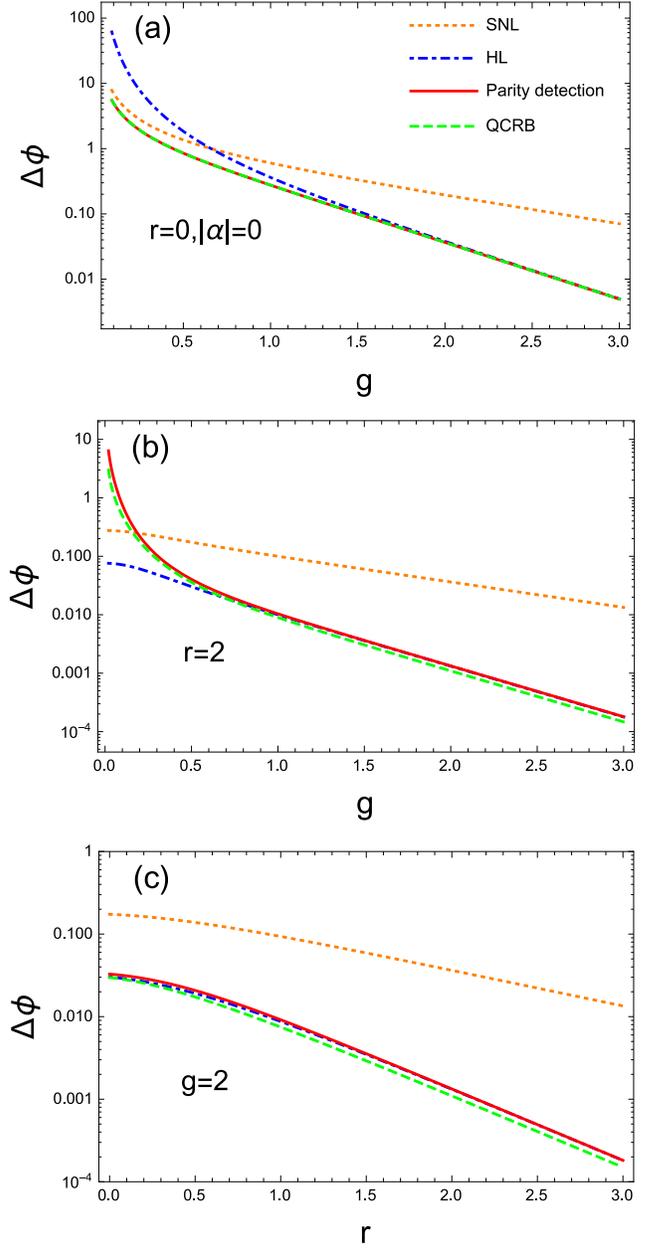}}\caption{(Color
online) Sensitivity of phase estimation with parity detection as a function of
(a) $g$ with vacuum inputs $r=0$ and $|\alpha_0|=0$, (b) $g$ with $r=2$ and $|\alpha_0|=\tanh(2g)e^{r}/2$, (c) $r$ with $g=2$ and $|\alpha_0|=\tanh(2g)e^{r}/2$. The dotted-orange line is for SNL, the dash-dotted-blue is for
HL, the dashed-green is for QCRB, and the solid-red is for sensitivity estimation based on parity detection.\label{fig2}}%
\end{figure}

When vacuum inputs are injected, the Heisenberg limit is found to be $\Delta\phi_{\text{HL}}=1/N_{\text{Tot}}=1/N_{\text{OPA}}$ while the corresponding phase sensitivity with parity detection is $\Delta\phi_{\text{p,V}}=1/\sqrt{N_{\text{OPA}}(N_{\text{OPA}}+2)}$. Fig. \ref{fig2}(a) compares the phase sensitivity $\Delta\phi_{\text{p,V}}$ with HL and SNL, as a function of OPA strength $g$, under the condition of vacuum input. It reveals that parity detection always beats HL. With the increase of $g$, HL becomes more and more close to phase sensitivity $\Delta\phi$. We notice that SNL is below HL when $g\lesssim0.6$ which is due to the total photon number $N_{\text{Tot}}<1$.

Next, we consider coherent and squeezed vacuum input states.
Letting $\Delta\phi \simeq \Delta\phi_{\text{HL}}$, according to Eq. (\ref{P}) and Eq. (\ref{HL}) and in the limit of $r\gg 1$ and $g\gg 1$, the optimal condition is found to be%
\begin{equation}
|\alpha_{0}|\simeq\frac{\tanh(2g)e^{r}}{2}.
\label{optimal}
\end{equation}
This expression reveals the requirement for the input coherent state $|\alpha_{0}|$, the input squeezed vacuum state $r$ and the OPA process $g$. We plot the phase sensitivity $\Delta\phi$ as a function of OPA strength $g$ in Fig. \ref{fig2}(b) which presents the comparison between  $\Delta\phi_{\text{p}}$ and $(\Delta\phi)_{\text{HL}}$. Under the condition $r=2$ and $|\alpha_{0}|=\tanh(2g)e^{r}/2$, the phase sensitivity approaches the HL when $g>2$ ($\tanh(2g)\simeq1$). Then Eq. (\ref{optimal}) is simplified to
$N_{\alpha}\simeq e^{2r}/4\simeq\sinh^{2}r=N_{s}$. The total photon number
simplifies to $N_{\text{Tot}}\simeq2N_{\text{OPA}}N_{\alpha}.$ Then the phase sensitivity
$\Delta\phi$ with parity detection always approaches the HL:%
\begin{equation}
\Delta\phi_{\text{p}}\simeq\frac{1}{\sqrt{4N_{\alpha}^{2}N_{\text{OPA}}^{2}}}\simeq\frac
{1}{N_{\text{Tot}}},
\end{equation}
When $g<1$, increasing the input squeezed parameter $r$ and increasing input coherent mean photon number enables the phase sensitivity to beat the SNL, but it does not approach HL. Fig. \ref{fig2}(c) is a plot of  the phase sensitivity $\Delta\phi$ as a function of input squeezed parameter $r$. Given $g=2$ and $|\alpha_{0}|=\tanh(2g)e^{r}/2$, the phase sensitivity is always below SNL and close to HL. 

When vacuum inputs $|\alpha_{0}|=0$ and $r=0$, parity detection always reaches below HL. With coherent and squeezed vacuum input $r\gg1$, the optimal condition of approaching HL is $|\alpha_{0}|\simeq e^{r}/2\simeq\sinh r$ and $\tanh(2g)\simeq 1$ which yields that the input coherent state and squeezed vacuum state should have approximately equal mean photon numbers. Thus the optimal condition for parity detection is exactly same as homodyne detection.

\subsection{Quantum Cram\'er-Rao Bound}

So far, we have shown that parity detection can achieve Heisenberg-limit
phase sensitivity in the SU(1,1) interferometer with coherent and squeezed
vacuum input states. In this section we will investigate quantum Cram\'er-Rao
bound of an SU(1,1) interferometer and compare the optimal phase sensitivity by parity detection with the QCRB which gives an
upper limit to the precision of quantum parameter estimation.

Recently, Gao \emph{et al}.~\cite{Gao} developed a general formalism for quantum
Cram\'er-Rao bound of Gaussian states, in which they can be fully expressed in
terms of the mean displacement and covariance matrix of the Gaussian state. In
this paper, we utilized this method to obtain the QCRB of our interferometric scheme.

The QCRB for an SU(1,1) interferometer with coherent and squeezed vacuum
inputs reads, see Appendix \ref{sec:bound} for details,

\begin{align}
\Delta\phi_{\text{QCRB}} =  &  \{N_{\text{OPA}}[N_{\text{OPA}}(2N_{s}+1)+2]\nonumber\\
&  \times(N_{s}+1)+2(N_{\text{OPA}}+2)N_{\alpha}\nonumber\\
&  \times\lbrack N_{\text{OPA}}(N_{s}+\sqrt{N_{s}(N_{s}+1)}+1)\nonumber\\
&  +1]\}^{-1/2},
\label{Qeq}
\end{align}
where the term on the right-hand side shows that $\Delta\phi_{\text{QCRB}}$ is related to not only input coherent intensity and input squeezed-vacuum intensity, but also optical parametric strength. When vacuum input is present, namely $N_s=0$ and $N_{\alpha}=0$, then QCRB is reduced to $\Delta
\phi_{\text{QCRB}}=1/\sqrt{N_{\text{OPA}}(N_{\text{OPA}}+2)}$. Thus parity detection saturates
QCRB with vacuum input as depicted in Fig. \ref{fig2}(a). Meanwhile, we notice that parity detection beats the HL. It is now well-known that states of indefinit total photon number can beat the HL \cite{Anisimov}. With coherent and squeezed vacuum input,  $\Delta\phi_{\text{QCRB}}<\Delta\phi_{\text{p}}$ always succeeds. As shown in Figs. \ref{fig2}(b) and \ref{fig2}(c), $\Delta\phi_{\text{QCRB}}$ is always below phase sensitivity with parity detection.
\begin{table*}[!h]
\tabcolsep 3mm
\doublerulesep 8mm
\caption{The quantum Cram\'er-Rao bound (QCRB) of an SU(1,1)
interferometer with different input states.}
\begin{center}
\renewcommand\arraystretch{2}
\begin{tabular}{p{0.15\textwidth}p{0.23\textwidth}p{0.351\textwidth}}
\hline
Input states & Parity detection & QCRB\\\hline
$|0\rangle\otimes|0\rangle$ & $1/\mathcal{K}^{1/2}$ \cite{Lij}  &  $1/\mathcal{K}^{1/2}$ \\
$|\alpha_{0}\rangle\otimes|0\rangle$ &$1/[\mathcal{K}(N_{\alpha}+1)]^{1/2}$ \cite{Lij} & $1/[2N_{\alpha}(\mathcal{K}+N_{\text{OPA}}+2)+\mathcal{K}]^{1/2}$\\
$|\frac{i\alpha_{0}}{\sqrt{2}}\rangle \otimes|\frac{\alpha_{0}}{\sqrt{2}}\rangle$ &  NEK & $1/\{2N_{\alpha}[\mathcal{K}+(N_{\text{OPA}}+1)\sqrt{\mathcal{K}}+1]+\mathcal{K}\}^{1/2}$\\
$|\alpha_{0}\rangle\otimes|0,\xi\rangle$ & $1/[\mathcal{K}(N_{\alpha}e^{2r}+\cosh^{2}r)]^{1/2}$ &  $1/\{\mathcal{K}[2N_{\alpha}(\sqrt{(N_{s}+1)N_{s}}+N_{s}+1)+(2N_{s}+1)(N_{s}+1)]+N_{\alpha}+2N_{s}(N_{s}+1)\}^{1/2} $ \\\hline
\end{tabular}\\
{where $\mathcal{K}=N_{\text{OPA}}(N_{\text{OPA}}+2)$ and NEK means not exactly known. Row 1: vacuum input;
Row 2: one-coherent input; Row 3: two-coherent input; Row 4: coherent mixed with squeezed-vacuum input.\label{Tab001}}
\end{center}

\end{table*}
\begin{table*}[ptbh]
\caption{The quantum Cram\'er-Rao bound (QCRB) of an MZI with different input states.}%
\tabcolsep 3mm \doublerulesep 8mm
\par
\begin{center}
\renewcommand\arraystretch{2}
\par
\begin{tabular}
[c]{lll}\hline
Input states & Parity detection & QCRB\\\hline
$|\alpha_{0}\rangle\otimes|0\rangle$ & $1/\sqrt{N_{\alpha}}$ & $1/\sqrt{N_{\alpha}}$\\
$|\frac{i\alpha_{0}}{\sqrt{2}}\rangle \otimes|\frac{\alpha_{0}}{\sqrt{2}}\rangle$ &  Always bad & $1/\sqrt{N_{\alpha}}$\\
$|\alpha_{0}\rangle\otimes|0,\xi\rangle$ & $1/\sqrt{N_{\alpha}e^{2r}+\sinh
^{2}r}$\cite{Pezze,Seshadreesan} & $1/\sqrt{N_{\alpha}e^{2r}+\sinh
^{2}r}$\cite{Pezze,Seshadreesan}\\\hline
\end{tabular}\\
{where we consider that the phase shifters are on both two arms, top arm with $\phi/2$\\ and bottom arm with $-\phi/2$. Row 1: one-coherent input; Row 2: two-coherent input;\\ Row 3: coherent mixed with squeezed-vacuum input.\label{Tab002}}
\end{center}
\end{table*}
\begin{table*}[ptbh]
\caption{The phase sensitivity for an SU(1,1)
interferometer with different detection methods.}%
\tabcolsep 3mm \doublerulesep 8mm
\par
\begin{center}
\renewcommand\arraystretch{2}
\par
\begin{tabular}
[c]{llll}\hline
Input states & Parity detection & Homodyne detection & Intensity detection\\\hline
$|0\rangle\otimes|0\rangle$ &$1/\mathcal{K}^{1/2}$ \cite{Lij} & Always bad & $1/\mathcal{K}^{1/2}$ \cite{Yurke,Marino}\\
$|\alpha_{0}\rangle\otimes|0\rangle$ & $1/[\mathcal{K}(N_{\alpha}+1)]^{1/2}$ \cite{Lij} & $1/(\mathcal{K}N_{\alpha})^{1/2}$ \cite{Li} &NEK\\
$|\frac{i\alpha_{0}}{\sqrt{2}}\rangle \otimes|\frac{\alpha_{0}}{\sqrt{2}}\rangle$ &  NEK & $\approx1/(2\mathcal{K}N_{\alpha})^{1/2}$ \cite{Li} &$1/(\mathcal{K}N_{\alpha})^{1/2}$ \cite{Plick}\\
$|\alpha_{0}\rangle\otimes|0,\xi\rangle$ & $1/[\mathcal{K}(N_{\alpha}e^{2r}+\cosh^{2}r)]^{1/2}$ &  $1/[\mathcal{K}N_{\alpha}e^{2r}]^{1/2}$ \cite{Li} & NEK\\\hline
\end{tabular}\\
{where NEK means not exactly known. Row 1: vacuum input;
Row 2: one-coherent input;\\ Row 3: two-coherent input; Row 4: coherent mixed with squeezed-vacuum input.\label{TabPHI}}
\end{center}
\end{table*}

In Table \ref{Tab001}, we present a comparison between phase sensitivity with parity detection and QCRB in an SU(1,1) interferometer with different inputs. Parity detection saturates QCRB only under the condition of vacuum input $|0\rangle\otimes|0\rangle$ while parity detection is always above QCRB with one-coherent and coherent mixed with squeezed vacuum input states. In the case of two-equal-coherent input states $|i\alpha_{0}/\sqrt{2}\rangle\otimes|\alpha_{0}/\sqrt{2}\rangle$, parity detection gives a worse minimum phase sensitivity where the analytical expression is not exactly known (NEK) due to divergence at the optimal phase point. We also found that two-equal-coherent input states have a lower QCRB than one coherent input state under the condition of same total input photon number. Table \ref{Tab002} shows the comparison of phase sensitivity with parity detection and QCRB in an SU(2) interferometer with different inputs. It reveals that although parity detection has poor statistics (due to parity detection signal $\langle\hat{\Pi}_{b}\rangle$ obtaining no information of phase shifter $\phi$) when two-equal-coherent-state input is used, parity detection achieves QCRB when one-coherent state input or coherent mixed with squeezed vacuum state input. Thus parity detection can be still considered as the optimal measurement scheme in an MZI. However, parity detection does not reach QCRB with coherent input or coherent mixed with squeezed vacuum in an SU(1,1) interferometer.
According to Tables \ref{Tab001} and \ref{Tab002}, the SU(1,1) interferometer has a better phase
sensitivity than the MZI by a roughly factor of $\mathcal{K}\equiv\sqrt{N_{\text{OPA}}(N_{\text{OPA}}+2)}$ with one coherent input $|\alpha_{0}\rangle\otimes|0\rangle$ and coherent mixed with squeezed-vacuum input $|\alpha_{0}\rangle\otimes|0,\xi\rangle$ due to amplification process. 

\section{Conclusion}

Table \ref{TabPHI} shows the comparison among parity detection, intensity detection, and homodyne detection for an SU(1,1)
interferometer with different input states. With vacuum inputs, parity measurement has
the same optimal phase sensitivity compared with intensity measurement \cite{Yurke,Marino}. Meanwhile the homodyne detection is always bad due to a constant measurement signal $\langle\hat{X}_2\rangle=0$. With one coherent input $|\alpha_{0}\rangle\otimes|0\rangle$ or coherent mixed with squeezed-vacuum input $|\alpha_{0}\rangle\otimes|0,\xi\rangle$, parity detection has a slightly better phase sensitivity than homodyne detection \cite{Li}, meanwhile intensity detection is not exactly known (NEK) due to divergence at the optimal point. In this point of view, parity detection is more suitable than homodyne detection. With two-coherent inputs, parity detection is NEK due to divergence at the optimal point, meanwhile homodyne detection is better than intensity detection. In this case, homodyne measurement is the optimal scheme.

In summary, we have investigated the parity detection on an SU(1,1)
interferometer with coherent and squeezed vacuum states as inputs. We have
presented that parity detection beats Heisenberg limit when coherent beam and
squeezed-vacuum beam are in roughly equal intensity with a stronger parameter
strength. Compared with homodyne detection, parity detection has a slightly
better optimal phase sensitivity with coherent and squeezed vacuum input
states. We have also shown a brief study of quantum Cram\'er-Rao bound of an
SU(1,1) interferometer. However, parity detection does not exactly reach the quantum
Cram\'er-Rao bound with a input. This motivates us to look for new optimal detection schemes to approach quantum Cram\'er-Rao bound in future work.

\section{Acknowledgement}

This work is supported by the National Basic Research Program of China (973
Program) under grant no. 2011CB921604, the National Natural Science
Foundation of China under grant nos 11474095, 11274118, 11234003, 11129402, and the fundamental research
funds for the central universities. DL is supported by the China Scholarship Council. BTG would like to acknowledge support from the Nation Physical Science Consortium and the Nation Institute of Standards \& Technology graduate fellowship program. JPD acknowledges support from the US National Science Foundation.

\appendix

\section{Parity Detection Signal}

\label{sec:signal} From Eqs. (\ref{wigner}) and (\ref{signal}), the
measurement signal $\langle\hat{\Pi}_{b}\rangle$ is found to be%
\begin{equation}
\langle\hat{\Pi}_{b}\rangle=\frac{1}{\sqrt{T_{1}}}e^{-T_{2}/T_{3}},
\end{equation}
where $T_{1} = e^{-2r} (e^{2r}+1)^{2} [8 \sinh^{4}(2g) (\cos(2\phi)-\cos \phi) + 4\cosh (4g) + 3 \cosh(8g) -7 ] + 64$, $T_{2}=4|\alpha|^{2}\sinh^{2}(2g)\{8\cosh
(4g)\cos(2\theta)\sin^{4}(  \phi/2)  -8\cosh(2g)\sin(2\theta
)\sin\phi(\cos\phi-1)+8e^{4r}[\cos\theta\sin\phi-2\cosh(2g)\sin\theta\sin
^{2}(  \phi/2)  ]^{2}+32e^{2r}\sinh^{2}(2g)\sin^{4}(
\phi/2)  +8\cosh(4g)\sin^{4}(  \phi/2)  -8\cos^{2}\theta
\cos\phi+[3\cos(2\theta)-1]\cos(2\phi)+\cos(2\theta)+5\}$, and $T_{3} = (e^{2r}+1)^{2}[8 \cosh(8g)\sin^{4}(\phi/2) + 8\cosh(4g)\sin^{2} \phi + 4 \cos \phi + 3 \cos (2 \phi) - 7] + 64 e^{2r}$. Letting $\phi=0,$ we find that signal is reduced to
\begin{equation}
\langle\hat{\Pi}_{b}\rangle|_{\phi=0}=1,
\end{equation}
which matches our prediction. When $\phi=0,$ the second OPA would undo what
the first one does causing the output fields to be the same as the inputs. Thus the
output in mode $b$ is the one-mode squeezed vacuum. For the one-mode squeezed
vacuum, parity signal is $1$ due to only even number distribution in the Fock
basis with $|0,\xi=re^{i\phi_{s}}\rangle=\sqrt{1/\cosh r}\sum_{n=0}^{\infty
}(\sqrt{(2n)!}/n)(1/2)^{n}[\exp(i\phi_{s})\tanh r]^{n}|2n\rangle.$
\cite{Barnett}

\section{quantum Cram\'er-Rao bound}

\label{sec:bound} First we will focus on evolution of mean values and
covariance matrix of quadrature operators in an SU(1,1) interferometer. Second
we will transform from the quadrature operator basis to the annihilation (creation)
operator basis. Then according to Ref.~\cite{Gao}, the QCRB will be obtained by mean
values and covariance matrix of annihilation (creation) operators.

$\hat{a}_{i}$ $(\hat{a}_{i}^{\dag})$ and $\hat{b}_{i}$ $(\hat{b}_{i}^{\dag}%
)$\ are the annihilation (creation) operators as shown in Fig. \ref{fig1}. We
introduce the quadrature operators $\hat{x}_{a_{i}}=\hat{a}_{i}+\hat{a}%
_{i}^{\dag},$ $\hat{p}_{a_{i}}=-i(\hat{a}_{i}-\hat{a}_{i}^{\dag}),$ $\hat
{x}_{b_{i}}=\hat{b}_{i}+\hat{b}_{i}^{\dag}$ and $\hat{p}_{b_{i}}=-i(\hat
{b}_{i}-\hat{b}_{i}^{\dag})$. We define quadrature column vector
$\mathbf{X}_{i}=(\hat{X}_{i,1},\hat{X}_{i,2},\hat{X}_{i,3},\hat{X}%
_{i,4})^{\intercal}\equiv(\hat{x}_{a_{i}},\hat{p}_{a_{i}},\hat{x}_{b_{i}%
},\hat{p}_{b_{i}})^{\intercal}$.

Next, we focus on the column vector of expectation values of the quadratures
$\mathbf{\bar{X}}_{i}$ and the symmetrized covariance matrix $\Gamma_{i}$
\cite{Gao,Braunstein2005,Weedbrook,Monras} where
\begin{equation}
\mathbf{\bar{X}}_{i}=(\langle\hat{X}_{i,1}\rangle,\langle\hat{X}_{i,2}%
\rangle,\langle\hat{X}_{i,3}\rangle,\langle\hat{X}_{i,4}%
\rangle)^{\intercal},
\end{equation}%
\begin{equation}
\Gamma_{i}^{kl}=\frac{1}{2}\text{Tr}[(\tilde{X}_{i,k}\tilde{X}_{i,l}+\tilde
{X}_{i,l}\tilde{X}_{i,k})\rho],
\end{equation}
with $\tilde{X}_{i,k}=\hat{X}_{i,k}-\langle\hat{X}_{i,k}\rangle
,\tilde{X}_{i,l}=\hat{X}_{i,l}-\langle\hat{X}_{i,l}\rangle$ and $\rho$
the density matrix.

The input-output relation of $\mathbf{\bar{X}}_{i}$ and $\Gamma_{i}$ can be
described as \cite{Adesso}%

\begin{equation}
\mathbf{\bar{X}}_{2}=S\mathbf{\bar{X}}_{0}, \label{x}%
\end{equation}%
\begin{equation}
\Gamma_{2}=S\Gamma_{0}S^{\intercal}, \label{gama}%
\end{equation}
where $\mathbf{\bar{X}}_{2}$ $(\mathbf{\bar{X}}_{0})$ and $\Gamma_{2}$
$(\Gamma_{0})$ are column vector of expectation values of the quadratures and
symmetrized covariance matrix for the output (input) states, respectively, $S$
is the transformation matrix. In general, transformation through the first
OPA, phase shifter and the second OPA could be given by %
\begin{equation}
S_{\text{OPA1}}=\left(
\begin{array}
[c]{cccc}%
\cosh g & 0 & \sinh g & 0\\
0 & \cosh g & 0 & -\sinh g\\
\sinh g & 0 & \cosh g & 0\\
0 & -\sinh g & 0 & \cosh g
\end{array}
\right)  ,
\end{equation}%
\begin{equation}
S_{\phi}=\left(
\begin{array}
[c]{cccc}%
\cos\phi & -\sin\phi & 0 & 0\\
\sin\phi & \cos\phi & 0 & 0\\
0 & 0 & 1 & 0\\
0 & 0 & 0 & 1
\end{array}
\right)  ,
\end{equation}%
\begin{equation}
S_{\text{OPA2}}=\left(
\begin{array}
[c]{cccc}%
\cosh g & 0 & -\sinh g & 0\\
0 & \cosh g & 0 & \sinh g\\
-\sinh g & 0 & \cosh g & 0\\
0 & \sinh g & 0 & \cosh g
\end{array}
\right)  ,
\end{equation}
where we have considered the balanced situation that $\theta_{1}=0,$
$\theta_{2}=\pi$ and $g_{1}=g_{2}=g.$ Therefore, the matrix can be
obtained as $S=S_{\text{OPA2}}S_{\phi}S_{\text{OPA1}}.$

In our case of a coherent and squeezed vacuum input states ($|\alpha_{0}%
\rangle\otimes|0,\xi=re^{i\phi_{s}}\rangle$), the initial mean values of
quadratures $\mathbf{\bar{X}}_{0}$ and covariance matrix $\Gamma_{0}$ are%
\begin{equation}
\mathbf{\bar{X}}_{0}=\left(
\begin{array}
[c]{cccc}%
2|\alpha_{0}| & 0 & 0 & 0
\end{array}
\right)  ^{\intercal},
\end{equation}%
\begin{equation}
\Gamma_{0}=\left(
\begin{array}
[c]{cccc}%
1 & 0 & 0 & 0\\
0 & 1 & 0 & 0\\
0 & 0 & e^{2r} & 0\\
0 & 0 & 0 & e^{-2r}%
\end{array}
\right)  ,
\end{equation}
where we have let $\theta_{\alpha}=0$ and $\phi_{s}=0.$ According to Eqs.
(\ref{x}) and (\ref{gama}), the final states can be found to be%
\begin{equation}
\mathbf{\bar{X}}_{2}=2|\alpha_{0}|\left(
\begin{array}
[c]{c}%
\cosh^{2}g\cos\phi-\sinh^{2}g\\
\cosh^{2}g\sin\phi\\
\sinh g\cosh g(1-\cos\phi)\\
\sinh g\cosh g\sin\phi
\end{array}
\right)  , \label{x2}%
\end{equation}%
\begin{equation}
\Gamma_{2}=\left(
\begin{array}
[c]{cccc}%
\gamma_{11} & \gamma_{12} & \gamma_{13} & \gamma_{14}\\
\gamma_{21} & \gamma_{22} & \gamma_{23} & \gamma_{24}\\
\gamma_{31} & \gamma_{32} & \gamma_{33} & \gamma_{34}\\
\gamma_{41} & \gamma_{42} & \gamma_{43} & \gamma_{44}%
\end{array}
\right)  , \label{gama2}%
\end{equation}
where%

\begin{align}
\gamma_{11}=  &  e^{-2r}\{e^{2r}\cos^{2}\phi\cosh^{4}g+[e^{2r}\cosh^{2}%
r\sin^{2}\phi \nonumber\\
&  +(e^{4r}\cos^{2}\phi-2e^{2r}\left(  1+e^{2r}\right)  \cos\phi+e^{4r} \nonumber\\
&  +\sin^{2}\phi)\sinh^{2}g]\cosh^{2}g+e^{2r}\sinh^{4}g\},
\end{align}

\begin{align}
\gamma_{12}=  &  e^{-2r}\cosh g\cosh r\sin\phi\nonumber\\
&  \times\lbrack e^{2r}\cos\phi\cosh^{2}g-e^{2r}\cos\phi\cosh^{2}r\nonumber\\
&  +\left(  -1+e^{4r}\right)  (\cos\phi-1)\sinh^{2}g]\nonumber\\
=& \gamma_{21},
\end{align}%
\begin{align}
\gamma_{13}=  &  e^{-2r}\cosh g\sinh g\nonumber\\
&  \times\{e^{2r}\left(  e^{2r}-\cos\phi\right)  (\cos\phi-1)\cosh
^{2}g\nonumber\\
&  -e^{2r}\cosh^{2}r\sin^{2}\phi+2\left(  1+e^{2r}\right) \nonumber\\
&  \times\lbrack\left(  -1+e^{2r}\right)  \cos\phi-1]\sin^{2}\frac{\phi}%
{2}\sinh^{2}g\}\nonumber\\
=& \gamma_{31},
\end{align}%
\begin{align}
\gamma_{14}=  &  e^{-2r}\cosh g\sin\phi\sinh g\nonumber\\
&  \times\{e^{2r}\cos\phi\cosh^{2}g+(1-e^{2r}\cos\phi\nonumber\\
&  +e^{2r})\cosh^{2}r+\left(  1+e^{2r}\right) \nonumber\\
&  \times\lbrack\left(  -1+e^{2r}\right)  \cos\phi-e^{2r}]\sinh^{2}g\}\nonumber\\
=& \gamma_{41},
\end{align}%
\begin{align}
\gamma_{22}=  &  e^{-2r}\{e^{2r}\cos^{2}\phi\cosh^{4}r\nonumber\\
&  +[e^{2r}\cosh^{2}g\sin^{2}\phi+(\cos^{2}\phi\nonumber\\
&  -2\left(  1+e^{2r}\right)  \cos\phi+e^{4r}\sin^{2}\phi+1)\nonumber\\
&  \times\sinh^{2}g]\cosh^{2}r+e^{2r}\sinh^{4}g\},
\end{align}%
\begin{align}
\gamma_{23}=  &  e^{-2r}\cosh r\sin\phi\sinh g\nonumber\\
&  \times\{e^{2r}(-\cos\phi+e^{2r}+1)\cosh^{2}g\nonumber\\
&  +e^{2r}\cos\phi\cosh^{2}r-(1+e^{2r})\nonumber\\
&  \times\lbrack(-1+e^{2r})\cos\phi+1]\sinh^{2}g\}\nonumber\\
=& \gamma_{32},
\end{align}%
\begin{align}
\gamma_{24}=  &  e^{-2r}\cosh r\sinh g\{(\cos\phi-1)\nonumber\\
&  \times(e^{2r}\cos\phi-1)\cosh^{2}r\nonumber\\
&  +e^{2r}\cosh^{2}g\sin^{2}\phi\nonumber\\
&  +2(1+e^{2r})[(-1+e^{2r})\cos\phi\nonumber\\
&  +e^{2r}]\sin^{2}\frac{\phi}{2}\sinh^{2}g\}\nonumber\\
=& \gamma_{42},
\end{align}%
\begin{align}
\gamma_{33}=  &  e^{-2r}\{e^{4r}\cosh^{4}g-e^{2r}[-\cos^{2}\phi\nonumber\\
&  +2(1+e^{2r})\cos\phi-1]\sinh^{2}g\cosh^{2}g\nonumber\\
&  +(e^{4r}\cos^{2}\phi+\sin^{2}\phi)\sinh^{4}g\nonumber\\
&  +e^{2r}\cosh^{2}r\sin^{2}\phi\sinh^{2}g\},
\end{align}%
\begin{align}
\gamma_{34}=  &  -e^{-2r}\sin\phi\sinh^{2}g[-e^{2r}\nonumber\\
&  \times(-\cos\phi+e^{2r}+1)\cosh^{2}g\nonumber\\
&  +(-e^{2r}\cos\phi+e^{2r}+1)\cosh^{2}r\nonumber\\
&  +(-1+e^{4r})\cos\phi\sinh^{2}g]\nonumber\\
=& \gamma_{43},
\end{align}%
and%
\begin{align}
\gamma_{44}=  &  e^{-2r}\{\cosh^{4}r+[e^{2r}\cos^{2}\phi-2(1+e^{2r}%
)\nonumber\\
&  \times\cos\phi+e^{2r}]\sinh^{2}g\cosh^{2}r\nonumber\\
&  +(\cos^{2}\phi+e^{4r}\sin^{2}\phi)\sinh^{4}g\nonumber\\
&  +e^{2r}\cosh^{2}g\sin^{2}\phi\sinh^{2}g\}.
\end{align}

So far, we have obtained the output quadrature vector and its covariance
matrix. Next, we will calculate the corresponding creation and annihilation
operator vector $\mathbf{d=(}d_{1},d_{2},d_{3},d_{4}\mathbf{)}^{\intercal
}\equiv(\hat{a}_{2},\hat{a}_{2}^{\dag},\hat{b}_{2},\hat{b}_{2}^{\dag
})^{\intercal}$ and its covariance matrix $\Sigma$ where matrix elements
$\Sigma^{u,v}=(1/2)\text{Tr}[\rho(\tilde{d}_{u}\tilde{d}_{v}+\tilde{d}_{v}\tilde
{d}_{u})]$ with $\tilde{d}_{u}=d_{u}-\bar{d}_{u}$ in terms of $\bar{d}%
_{u}=\text{Tr}[\rho d_{u}].$ The commutation relations are described as $[d_{u}%
,d_{v}]=\Omega^{u,v},$ where
\begin{equation}
\Omega=\left(
\begin{array}
[c]{cccc}%
0 & 1 & 0 & 0\\
-1 & 0 & 0 & 0\\
0 & 0 & 0 & 1\\
0 & 0 & -1 & 0
\end{array}
\right)  .
\end{equation}
Equivalently, the relations between $\mathbf{\bar{d}}$ ($\Sigma$) and
$\mathbf{\bar{X}}_{2}$ ($\Gamma_{2}$) are expressed as%
\begin{equation}
\mathbf{\bar{d}}=H\mathbf{\bar{X}}_{2},\label{d}%
\end{equation}%
\begin{equation}
\Sigma=\frac{1}{2}H\Gamma_{2}H^{\intercal},\label{sigama}%
\end{equation}
where $\mathbf{\bar{d}=(}\bar{d}_{1},\bar{d}_{2},\bar{d}_{3},\bar{d}%
_{4}\mathbf{)}^{\intercal}$ is mean value of $\mathbf{d}$ and%
\begin{equation}
H=\frac{1}{\sqrt{2}}\left(
\begin{array}
[c]{cccc}%
1 & i & 0 & 0\\
1 & -i & 0 & 0\\
0 & 0 & 1 & i\\
0 & 0 & 1 & -i
\end{array}
\right)  .
\end{equation}
According to Ref. \cite{Gao}, the quantum Fisher information is given by%
\begin{align}
F= &  \frac{1}{2}\text{Tr}\{\partial_{\phi}\Sigma\lbrack\Sigma(\partial_{\phi}%
\Sigma)^{-1}\Sigma^{\intercal}+\frac{1}{4}\Omega(\partial_{\phi}\Sigma
)^{-1}\Omega^{\intercal}]^{-1}\}\nonumber\\
&  +(\partial_{\phi}\mathbf{\bar{d})}^{\intercal}(\Sigma)^{-1}(\partial_{\phi
}\mathbf{\bar{d}),}\label{fisher}%
\end{align}
where $\partial_{\phi}\Sigma=\partial\Sigma/\partial\phi$ and $\partial_{\phi
}\mathbf{\bar{d}=\partial\bar{d}}/\partial\phi$. Then the corresponding
quantum Cram\'er-Rao bound is given by%
\begin{equation}
\Delta\phi_{\text{QCRB}}=\frac{1}{\sqrt{F}}.\label{QCRB}%
\end{equation}
Combined with Eqs. (\ref{x2}), (\ref{gama2}), (\ref{d}), (\ref{sigama}),
(\ref{fisher}), and (\ref{QCRB}), the quantum Cram\'er-Rao bound is found to be%
\begin{align}
\Delta\phi_{\text{QCRB}} &  =\{2(N_{\text{OPA}}+2)[N_{\text{OPA}}(\cosh^{2}r\nonumber\\
&  +\cosh r\sinh r)+1]N_{\alpha}\nonumber\\
&  +N_{\text{OPA}}[N_{\text{OPA}}(\cosh^{2}r+\sinh^{2}r)\nonumber\\
&  +2]\cosh^{2}r\}^{-1/2}.\label{AQ}%
\end{align}
Inserting $\sinh^{2}r\equiv N_{s}$ into Eq. (\ref{AQ}), then QCRB is described
by%
\begin{align}
\Delta\phi_{\text{QCRB}}= &  \{N_{\text{OPA}}[N_{\text{OPA}}(2N_{s}+1)+2]\nonumber\\
&  \times(N_{s}+1)+2(N_{\text{OPA}}+2)N_{\alpha}\nonumber\\
&  \times\lbrack N_{\text{OPA}}(N_{s}+\sqrt{N_{s}(N_{s}+1)}+1)\nonumber\\
&  +1]\}^{-1/2}.
\end{align}

\end{document}